\documentclass[runningheads]{llncs}

\usepackage[T1]{fontenc}
\usepackage{graphicx}
\usepackage{amsmath,amssymb}
\usepackage{booktabs}
\usepackage{tabularx}
\usepackage{array}
\usepackage{makecell}
\usepackage{multirow}
\usepackage{longtable}
\usepackage{placeins}
\usepackage{needspace}
\usepackage[hidelinks]{hyperref}

\graphicspath{{figures/}}

\begin{document}

\title{Simulation-Based Quantum Forensics for QKD: Machine Learning Attack Detection and Signature Extraction}

\author{
Madhab Chandra Das\inst{1}\orcidID{0000-0003-2724-4315}
\thanks{Corresponding author: Madhab Chandra Das,
\texttt{mxd147@shsu.edu}.}
\and
Sajid Ahmed\inst{1}\orcidID{0009-0008-0003-5429}
\and
Uzair Hussain\inst{1}\orcidID{0009-0005-9019-7242}
\and
Van Vung Pham\inst{1}\orcidID{0000-0001-9702-8904}
}

\institute{
Department of Computer Science, Sam Houston State University,
Huntsville, TX, USA\\
\email{
mxd147@shsu.edu,
sxa202@shsu.edu,
uxh002@shsu.edu,
vung.pham@shsu.edu
}
}

\maketitle
\pagestyle{empty}
\thispagestyle{empty}

\begin{abstract}

Quantum Key Distribution (QKD) supports secure key establishment under ideal protocol assumptions, but practical systems remain exposed to implementation-level attacks against optical channels, detectors, and controller behavior. This study presents Quantum Forensic Log (QF-LOG), a simulation-based forensic framework for analyzing aggregated QKD operational logs rather than raw quantum states. QF-LOG uses four observable features: quantum bit error rate (QBER), photon count, latency in milliseconds, and abort flag. The framework classifies normal operation and three simulated attack-related conditions: partial intercept-resend, detector-blind, and fiber-tap. A physics-inspired Generator-A dataset with 5 million balanced records was constructed using explicit mathematical generation models for all four features. A transparent threshold baseline, logistic regression, random forest, and XGBoost were evaluated on Generator-A. The full-feature learning models achieved macro F1-scores from 0.8211 to 0.8243 and outperformed the transparent two-feature comparator. Feature diagnostics showed that QBER contributed most, followed by latency in milliseconds and photon count, while no single feature reproduced the full feature set's performance. Robustness was assessed on a separately generated Generator-B dataset of 5 million records that uses the same governing equations under shifted simulated operating parameters. Without retraining or tuning, their macro F1-scores decreased to values from 0.5645 to 0.5669. These findings show that QF-LOG can support interpretable forensic classification in a controlled synthetic setting, but stable internal evaluation does not establish reliability under changed simulated QKD operating conditions. The framework provides a documented basis for future evaluation across broader simulated settings and hardware-generated QKD logs.

\end{abstract}


\section{Introduction}
\label{sec:introduction}

Quantum Key Distribution (QKD) allows two parties to establish a secret key through quantum signal transmission and authenticated classical communication \cite{gisin2002quantum,lo2014secure}. In BB84, interaction with transmitted quantum states can introduce observable errors during key establishment \cite{bennett1984bb84}. However, protocol-level security does not remove implementation weaknesses. Time-shift and detector-blinding attacks show that practical QKD analysis must also consider detector behavior and supporting device interfaces \cite{qi2007timeshift,lydersen2010hacking,scarani2009security}.

This study addresses a forensic question for QKD operation: after abnormal behavior is observed, which operational records can support structured investigation? We define the Quantum Forensic Log (QF-LOG) as an aggregated operational-log representation based on four observable features: QBER, photon count, latency in milliseconds, and abort flag. QF-LOG does not contain raw quantum states and is not treated as proof of a hardware attack. It is used to examine abnormal operational patterns under controlled simulation settings.

The framework considers one normal condition and three simulated attack-related conditions: partial intercept-resend, detector-blind, and fiber-tap. Partial intercept-resend increases QBER through a measured and re-prepared pulse fraction. Fiber-tap reduces observed photon counts through added optical attenuation. Detector-blind is modeled as a detector-blind suppression proxy motivated by documented detector-blinding attacks \cite{lydersen2010hacking}. These are simulation definitions and are not presented as observations from deployed QKD hardware.

The study uses two synthetic datasets. Generator-A supports model development, internal evaluation, repeated validation, and feature diagnostics. Generator-B is generated separately for cross-generator assessment. It keeps the same features and classes but changes simulated operating conditions. Methods fitted on Generator-A are tested directly on Generator-B without retraining or tuning.

The contributions of this study are:

\begin{itemize}
\item An aggregated operational-log representation for post-event analysis of simulated QKD behavior.
\item A fully documented equation-driven generation process for four observable QKD log attributes.
\item A comparison between a transparent two-feature rule comparator and full-feature learning models.
\item A controlled generator-shift assessment that exposes sensitivity to changed simulated channel and detector settings.
\item\ An explicit analysis of the limits of interpreting generator-embedded feature patterns as attack evidence.
\end{itemize}

The results show that the full-feature learning models outperform the transparent two-feature comparator on Generator-A, while no single stored feature reproduces full-feature performance. However, all classifiers degrade substantially on Generator-B. This finding shows that stable internal performance on one synthetic generator does not establish reliability under changed simulated QKD operating conditions.


\section{Related Work}
\label{sec:related}

Quantum Key Distribution (QKD) establishes secret keys through quantum signal transmission and authenticated classical communication. Foundational protocols such as BB84 and entanglement-based QKD showed that eavesdropping can introduce observable disturbances in quantum transmission \cite{bennett1984bb84}. Later surveys and practical studies showed that real QKD systems also face implementation-level risks beyond the ideal protocol model \cite{gisin2002quantum,scarani2009security,diamanti2016practical,pirandola2020advances}.

Prior work has demonstrated several practical attacks against QKD devices and channels, including detector-efficiency mismatch, time-shift, phase-remapping, wavelength-dependent, and detector-blinding attacks \cite{makarov2006,qi2007timeshift,gerhardt2011blinding,lydersen2010hacking}. These attacks exploit detector behavior, optical components, or control interfaces rather than breaking the underlying quantum protocol. Some attacks may also affect observable operational statistics without producing a simple or immediate QBER-only signature.

Several defenses have been proposed for practical QKD systems. Decoy-state methods address photon-number-splitting related weaknesses, while measurement-device-independent QKD reduces detector-side trust assumptions \cite{hwang2003decoy,lo2012mdi}. Detector monitoring and field studies have also shown the importance of observing QKD system behavior during operation \cite{yuan2011monitoring}. These works mainly focus on prevention, secure implementation, or operational demonstration. They do not directly study forensic interpretation of aggregated QKD operational logs after abnormal behavior is observed.

Machine learning and anomaly detection are widely used in security monitoring, but prior QKD studies mainly address protocol security, implementation attacks, or preventive monitoring. QF-LOG instead examines whether four aggregated operational-log features can support post-event classification and forensic interpretation under controlled simulation. Its contribution is the documented forensic representation and cross-generator evaluation, rather than a new classifier or hardware-validated attack detector.


\section{Threat Model and System Overview}
\label{sec:threatmodel}

The threat model considers abnormal behavior affecting the quantum channel between Alice and Bob or Bob's detector interface. The framework does not inspect raw quantum states. It examines aggregated QF-LOG observations that may reflect changes in transmission errors, detected signal level, processing delay, or protocol interruption.

For channel-side behavior, partial measurement and re-preparation is modeled as an error-related disturbance, while signal diversion is modeled as additional optical attenuation. For detector-side behavior, the framework uses a detector-blind suppression proxy motivated by detector-blinding attacks against practical QKD systems \cite{lydersen2010hacking}. These modeled conditions are used only for controlled simulation and do not represent observed attacks on deployed QKD hardware.

\begin{figure*}[t]
\centering
\includegraphics[width=\textwidth]{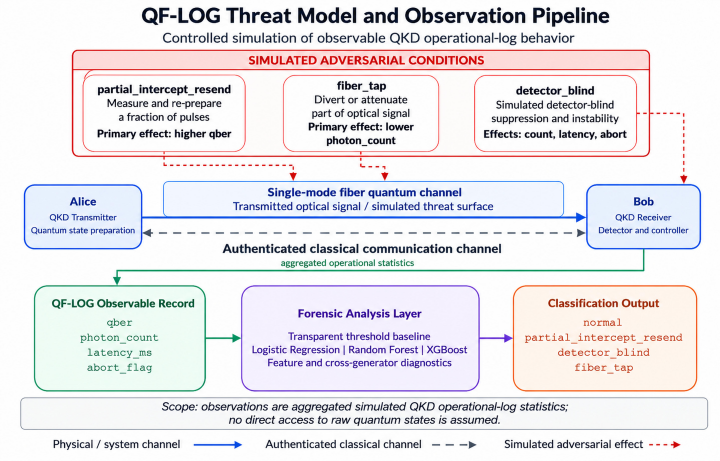}
\caption{System and threat model for the controlled QF-LOG simulation. 
The simulated adversarial conditions affect observable QKD operational-log features, which are analyzed through the forensic classification layer.}
\label{fig:threat}
\end{figure*}

Figure~\ref{fig:threat} summarizes the system boundary. Alice and Bob form the QKD endpoints, while the simulated conditions affect the observable operational-log features. The resulting QF-LOG records are analyzed by the forensic classification layer to distinguish the four simulated classes. Evaluation with hardware-generated QKD logs remains future work.

\section{Dataset Construction}
\label{sec:dataset_construction}

This study uses two equation-driven synthetic QF-LOG datasets under controlled simulation settings. Generator-A is used for model development, internal evaluation, and feature analysis. Generator-B is generated separately and used only for cross-generator assessment under changed simulated operating conditions. Both generators produce aggregated operational records; neither is presented as measured QKD hardware data.

\subsection{Generator-A Configuration}
\label{subsec:generator_a_configuration}

Generator-A contains 5,000,000 heterogeneous frame-level records, with 1,250,000 per simulated condition and N=16,000 transmitted pulses per frame. It does not represent a temporal trace from one fixed physical link. The classifiers use only four QF-LOG attributes: \texttt{qber}, \texttt{photon\_count}, \texttt{latency\_ms}, and \texttt{abort\_flag}. Source, channel, detector, and intervention variables are retained only as generation metadata. The stored labels are \texttt{normal}, \texttt{partial\_\allowbreak intercept\_\allowbreak resend}, \texttt{detector\_blind}, and \texttt{fiber\_tap}.

Table~\ref{tab:generator_a_parameters} summarizes the Generator-A settings. Attack-intervention quantities are active only for their corresponding simulated conditions. Thus, \(f_{\mathrm{IR}} = 0\) outside partial intercept-resend, \(a_{\mathrm{tap}} = 0\) outside fiber-tap, and \(r_{\mathrm{DB}} = 1\) with \(\delta_{\mathrm{DB}} = 0\) outside detector-blind.

\begin{table}[t]
\centering
\caption{Configured Generator-A parameters for controlled QF-LOG generation.}
\label{tab:generator_a_parameters}
\scriptsize
\setlength{\tabcolsep}{3pt}
\renewcommand{\arraystretch}{1.02}
\begin{tabularx}{\linewidth}{@{}l
>{\raggedright\arraybackslash}X
>{\raggedright\arraybackslash}X@{}}
\toprule
\textbf{Symbol} & \textbf{Meaning} & \textbf{Configuration} \\
\midrule
\(N\) & Pulses per frame & \(16{,}000\) \\
\(\mu\) & Mean photon number per pulse & \(\mathcal{N}(0.50,0.04^2)\), bounded to \([0.35,0.65]\) \\
\(L\) & Fiber length in km & \(\mathcal{N}(10.0,2.50^2)\), bounded to \([3.0,20.0]\) \\
\(\alpha\) & Fiber attenuation in dB/km & \(\mathcal{N}(0.20,0.01^2)\), bounded to \([0.17,0.23]\) \\
\(\eta_d\) & Detector efficiency & \(\mathcal{N}(0.20,0.018^2)\), bounded to \([0.14,0.26]\) \\
\(Y_0\) & Background yield probability & \(\mathcal{N}(5.0{\times}10^{-5},(1.0{\times}10^{-5})^2)\), bounded to \([1.0{\times}10^{-5},9.0{\times}10^{-5}]\) \\
\(e_d\) & Optical misalignment error & \(\mathcal{N}(0.018,0.006^2)\), bounded to \([0.005,0.040]\) \\
\(f_{\mathrm{IR}}\) & Attacked-pulse fraction & \(0.03+0.42\,\mathrm{Beta}(1.5,2.8)\) \\
\(a_{\mathrm{tap}}\) & Added optical attenuation in dB & \(0.10+2.50\,\mathrm{Beta}(1.5,2.2)\) \\
\(r_{\mathrm{DB}}\) & detector-blind factor & \(0.05+0.90\,\mathrm{Beta}(1.4,2.2)\) \\
\(\delta_{\mathrm{DB}}\) & Added detector-blind error & \(0.025\,\mathrm{Beta}(1.2,3.0)\) \\
\(T_0\) & Baseline latency in ms & \(\mathcal{N}(4.55,0.20^2)\), bounded to \([4.00,5.10]\) \\
\(\epsilon_T\) & Latency noise in ms & \(\mathcal{N}(0,0.15^2)\) \\
\bottomrule
\end{tabularx}
\end{table}

The source, channel, and detector variables follow weak coherent pulse fiber-based QKD parameter structure \cite{Ma2005PracticalDecoy}. The mean attenuation of \(0.20\) dB/km is consistent with standard telecommunication fiber values reported in prior QKD studies \cite{Ma2005PracticalDecoy,gobby2004qkd}. All distributions, bounded ranges, latency coefficients, abort rules, intervention settings, and Generator-B shifts are controlled simulation assumptions. They are not presented as hardware-calibrated measurements. Generated figures use the stored feature names and labels for consistency with the implemented pipeline.

\subsection{Observable Feature Generation}
\label{subsec:feature_generation}

The optical-loss, click-probability, baseline-error, and binary-entropy structures follow weak coherent pulse QKD modeling principles \cite{Ma2005PracticalDecoy}. The attack-intervention terms, latency rule, and abort rule are configured simulation components.

\subsubsection{Photon Count}

For each frame, channel transmission, total detection efficiency, and click probability are computed as
\begin{equation}
\begin{aligned}
\eta_{\mathrm{ch},i}
&=
10^{-\frac{\alpha_i L_i+a_{\mathrm{tap},i}}{10}}, \\
\eta_i
&=
\eta_{d,i}\eta_{\mathrm{ch},i}r_{\mathrm{DB},i}, \\
p_{\mathrm{click},i}
&=
1-(1-Y_{0,i})\exp(-\mu_i\eta_i).
\end{aligned}
\label{eq:channel_click_model}
\end{equation}

The observed photon count is sampled as
\begin{equation}
\texttt{photon\_count}_i
\sim
\mathrm{Binomial}
\left(
N,p_{\mathrm{click},i}
\right).
\label{eq:photon_count_generation}
\end{equation}
Here, fiber-tap changes photon count through added attenuation, and detector-blind changes it through detector-blind suppression.

\subsubsection{QBER}

The signal click probability and baseline error probability are
\begin{equation}
\begin{aligned}
p_{\mathrm{sig},i}
&=
1-\exp(-\mu_i\eta_i), \\
q_{\mathrm{base},i}
&=
\frac{0.5Y_{0,i}+e_{d,i}p_{\mathrm{sig},i}}
     {p_{\mathrm{click},i}}.
\end{aligned}
\label{eq:baseline_qber}
\end{equation}

The class-dependent error probability is
\begin{equation}
q_i^{*}
=
\begin{cases}
q_{\mathrm{base},i},
& \text{normal or fiber\_tap}, \\[3pt]
(1-f_{\mathrm{IR},i})q_{\mathrm{base},i}
+0.25f_{\mathrm{IR},i},
& \text{partial\_intercept\_resend}, \\[3pt]
q_{\mathrm{base},i}+\delta_{\mathrm{DB},i},
& \text{detector\_blind}.
\end{cases}
\label{eq:class_qber_model}
\end{equation}

The \(0.25\) term represents the BB84 intercept-resend error contribution for attacked pulses under random basis selection \cite{gisin2002quantum}. The detector-blind term is a simulation proxy motivated by documented detector-blinding attacks \cite{lydersen2010hacking}.

The error count and QBER are generated as
\begin{equation}
\begin{aligned}
E_i
&\sim
\mathrm{Binomial}
\left(
\texttt{photon\_count}_i,q_i^{*}
\right), \\
\texttt{qber}_i
&=
\begin{cases}
\dfrac{E_i}{\texttt{photon\_count}_i},
& \texttt{photon\_count}_i > 0, \\[6pt]
0,
& \texttt{photon\_count}_i = 0.
\end{cases}
\end{aligned}
\label{eq:qber_generation}
\end{equation}

\subsubsection{Latency in Milliseconds}

The expected photon count without attack intervention and the observed deficiency ratio are
\begin{equation}
\begin{aligned}
\eta_{\mathrm{ch},0,i}
&=
10^{-\frac{\alpha_iL_i}{10}}, \\
C_{\mathrm{ref},i}
&=
N\left[
1-(1-Y_{0,i})
\exp(-\mu_i\eta_{d,i}\eta_{\mathrm{ch},0,i})
\right], \\
R_{\mathrm{loss},i}
&=
\max
\left(
0,
1-\frac{\texttt{photon\_count}_i}{C_{\mathrm{ref},i}}
\right).
\end{aligned}
\label{eq:reference_count_loss}
\end{equation}

Using
\begin{equation}
H_2(q)
=
-q\log_2(q)-(1-q)\log_2(1-q),
\label{eq:binary_entropy}
\end{equation}
the latency feature is generated as
\begin{equation}
\texttt{latency\_ms}_i
=
T_{0,i}
+
5.50H_2(\texttt{qber}_i)
+
2.80R_{\mathrm{loss},i}
+
\epsilon_{T,i}.
\label{eq:latency_generation}
\end{equation}

Equation~\eqref{eq:latency_generation} is a configured simulation rule. Its coefficients increase latency under higher QBER or stronger photon-count deficiency. They are not hardware timing measurements.

\subsubsection{Abort Flag}

The abort feature is generated from the observed QF-LOG values. The three probability components are
\begin{equation}
\begin{aligned}
p_{Q,i}
&=
\sigma\!\left[90(\texttt{qber}_i-0.105)\right], \\
p_{L,i}
&=
\sigma\!\left[18(R_{\mathrm{loss},i}-0.55)\right], \\
p_{T,i}
&=
\sigma\!\left[3(\texttt{latency\_ms}_i-8.00)\right],
\qquad
\sigma(z)=\frac{1}{1+\exp(-z)}.
\end{aligned}
\label{eq:abort_components}
\end{equation}

With \(p_0=0.0005\), the abort probability and binary feature are
\begin{equation}
\begin{aligned}
p_{\mathrm{abort},i}
&=
1-
(1-p_0)
(1-p_{Q,i})
(1-p_{L,i})
(1-p_{T,i}), \\
\texttt{abort\_flag}_i
&\sim
\mathrm{Bernoulli}
\left(
p_{\mathrm{abort},i}
\right).
\end{aligned}
\label{eq:abort_generation}
\end{equation}

Equation~\eqref{eq:abort_generation} is a configured simulation rule. Its thresholds and sigmoid coefficients produce probabilistic abort behavior from observed features rather than assigning aborts directly from class labels.

\subsection{Generator-A Summary}
\label{subsec:generator_a_summary}

Table~\ref{tab:generator_a_summary} reports the observed feature behavior in Generator-A, partial intercept-resend produces the largest increase in QBER while maintaining a photon-count level close to normal operation. Fiber-tap reduces photon count through the configured attenuation term. Detector-blind produces the largest count reduction and the highest abort rate through the configured response suppression and abort rules.

\begin{table}[t]
\centering
\caption{Observed Generator-A feature summary.}
\label{tab:generator_a_summary}
\footnotesize
\renewcommand{\arraystretch}{1.10}
\begin{tabularx}{\linewidth}{@{}
>{\raggedright\arraybackslash}X
rrrrr@{}}
\toprule
\textbf{Condition} &
\textbf{Samples} &
\textbf{QBER} &
\textbf{Count} &
\textbf{Latency} &
\textbf{Abort Rate} \\
\midrule
Normal                  
& 1.25M & 0.01843 & 984.78 & 5.30 & 0.00161 \\
partial\_intercept\_resend 
& 1.25M & 0.05927 & 984.58 & 6.34 & 0.08462 \\
detector\_blind           
& 1.25M & 0.02660 & 400.69 & 7.16 & 0.60948 \\
fiber\_tap                
& 1.25M & 0.01857 & 773.44 & 5.87 & 0.01503 \\
\bottomrule
\end{tabularx}
\end{table}

\subsection{Generator-B for Cross-Generator Assessment}
\label{subsec:generator_b}

Generator-B tests stability under controlled parameter shift within
the same generation family. Table~\ref{tab:generator_b_compact}
summarizes its three class-independent simulated operating profiles:
shifted standard, low-efficiency detector, and longer-fiber link.

For each affected quantity, the across-frame offset is generated,
\begin{equation}
\begin{aligned}
d_{s,0} &\sim \mathcal{N}(0,\sigma_{0}^{2}), \\
d_{s,t} &= 0.96d_{s,t-1}+\epsilon_{s,t}, \qquad t \geq 1, \\
\epsilon_{s,t} &\sim \mathcal{N}(0,\sigma_{\epsilon}^{2}),
\end{aligned}
\label{eq:generator_b_frame_variation}
\end{equation}
where \(s\) denotes a simulated session and \(t\) denotes the frame position. The offsets modify channel loss, detector efficiency, misalignment error, and baseline latency.

\begin{table}[t]
\centering
\caption{Compact Generator-B profile settings used for cross-generator assessment.}
\label{tab:generator_b_compact}
\footnotesize
\setlength{\tabcolsep}{3pt}
\renewcommand{\arraystretch}{1.04}
\begin{tabular}{lccc}
\toprule
Parameter & Shifted standard & Low-efficiency detector & Longer-fiber link \\
\midrule
Profile probability & 0.45 & 0.30 & 0.25 \\
Mean photon number, \(\mu\) & 0.49 & 0.48 & 0.50 \\
Fiber length, \(L\) (km) & 12.0 & 12.5 & 18.0 \\
Attenuation, \(\alpha\) (dB/km) & 0.205 & 0.208 & 0.210 \\
Detector efficiency, \(\eta_d\) & 0.190 & 0.166 & 0.185 \\
Misalignment error, \(e_d\) & 0.022 & 0.026 & 0.028 \\
Baseline latency, \(T_0\) (ms) & 4.72 & 4.88 & 5.00 \\
\bottomrule
\end{tabular}
\end{table}

Generator-B also shifts the simulated intervention intensities. For each affected session, an intervention value is sampled from a beta-based distribution, perturbed by frame-level Gaussian jitter, and restricted to its documented range. This procedure is applied to the partial intercept-resend fraction, fiber-tap attenuation, detector response factor, and added detector response error component.

The optical transmission, detector-click, QBER, latency, and abort-generation structures remain the same as Generator-A. The Generator-B profiles, offsets, and intervention shifts are controlled simulation assumptions. They are not presented as QKD hardware measurements. Complete Generator-B settings are retained with the reproducibility artifacts.

\subsection{Scope and Reproducibility}
\label{subsec:dataset_scope}

This study is limited to controlled simulation-based QF-LOG generation and evaluation. The generated records are not hardware measurements and are not presented as evidence of attack detection on deployed QKD systems. To support reproducibility, the Generator-A and Generator-B settings, generation and evaluation scripts, metadata, random seeds, and result files are archived on Zenodo \cite{das2025qkdforensicdataset}.


\section{Methodology}
\label{sec:methodology}

This study evaluates whether observable QKD operational logs can support multiclass forensic classification under controlled simulation conditions. The mathematical construction of the synthetic QF-LOG records is presented in Section~\ref{sec:dataset_construction}. This section describes the analytical workflow applied after data generation. Generator-A is used for model development and internal evaluation, while the separately generated shifted Generator-B dataset is reserved for cross-generator assessment.


\subsection{Evaluation Design}
\label{subsec:evaluation_design}

Generator-A contains 5,000,000 balanced records. It is divided using a stratified 75:25 holdout split with random seed 42, producing 3,750,000 training records and 1,250,000 test records. All preprocessing parameters, baseline cut-points, and model parameters are estimated from the training partition only.

The evaluation compares a transparent rule-based baseline with logistic regression, random forest, and XGBoost. Accuracy, macro F1-score, and Matthews correlation coefficient (MCC) are reported for all methods. For the learning-based classifiers, macro ROC-AUC and macro PR-AUC are also reported from multiclass probability outputs.

Internal stability is examined through stratified five-fold validation repeated with seeds $\{1,7,21,42,99\}$, resulting in 25 evaluation runs per method. Mean and standard deviation are reported across these runs.

\subsection{Evaluated Methods}
\label{subsec:evaluated_methods}

The transparent baseline is a data-derived comparator based on QBER and photon-count behavior in the Generator-A training partition. Let \(\tilde{q}_k\) and \(\tilde{c}_k\) denote the class median QBER and photon count for class \(k\), where \(\mathrm{N}\), \(\mathrm{IR}\), \(\mathrm{DB}\), and \(\mathrm{FT}\) denote normal, partial intercept-resend, detector-blind, and fiber-tap. The cut points are

\begin{equation}
\tau_{\mathrm{DB}}=\frac{\tilde{c}_{\mathrm{DB}}+\tilde{c}_{\mathrm{FT}}}{2}, \quad
\tau_{\mathrm{IR}}=\frac{\tilde{q}_{\mathrm{N}}+\tilde{q}_{\mathrm{IR}}}{2}, \quad
\tau_{\mathrm{FT}}=\frac{\tilde{c}_{\mathrm{FT}}+\tilde{c}_{\mathrm{N}}}{2}.
\end{equation}

The sequential prediction rule is

\begin{equation}
\widehat{y} =
\begin{cases}
\mathrm{DB}, & c \leq \tau_{\mathrm{DB}}, \\
\mathrm{IR}, & q \geq \tau_{\mathrm{IR}}, \\
\mathrm{FT}, & c \leq \tau_{\mathrm{FT}}, \\
\mathrm{N},  & \text{otherwise},
\end{cases}
\label{eq:threshold_baseline}
\end{equation}
where \(q\) and \(c\) denote the QBER and photon-count values of an input record. Since the cut points are estimated from labeled training records, the baseline is a transparent two-feature comparator rather than a feature-matched alternative to the four-feature learning models or an operational decision rule.

The learning-based methods are logistic regression, random forest \cite{breiman2001random}, and XGBoost \cite{chen2016pxgboost}. Logistic regression uses training-data standardization, the \texttt{L-BFGS} solver, and \(1000\) maximum iterations. Random forest uses \(300\) trees, maximum depth \(14\), the Gini criterion, and bootstrap sampling. XGBoost uses \(400\) estimators, maximum depth \(10\), learning rate \(0.08\), and subsample and column-subsample ratios of \(0.90\). All applicable fixed-holdout random seeds are set to \(42\).

\subsection{Diagnostic and Statistical Analysis}
\label{subsec:diagnostic_analysis}

Feature contribution is examined using leave-one-feature-out ablation, single-feature evaluation, and permutation importance. In ablation, each learning-based classifier is retrained after removing one observable attribute, and the macro F1-score change is measured. Single-feature evaluation tests whether one attribute can reproduce full-feature performance. Permutation importance is computed on a balanced, unseen Generator-A diagnostic subset of 100,000 records, with each attribute permuted ten times and scored by macro F1 decrease.

Paired method comparisons use McNemar's test on predictions from the same held-out records \cite{mcnemar1947note}. Each learning-based classifier is compared with the transparent baseline, and random forest is compared with XGBoost because their scores are close. The same procedure is applied to Generator-B. Statistical results are interpreted with macro F1 differences to avoid overstating small effects.

\subsection{Cross-Generator Assessment and Forensic Interpretation}
\label{subsec:cross_generator_assessment}

Methods fitted on Generator-A are applied directly to Generator-B without retraining, recalibration, or parameter tuning. Generator-B preserves the same features and class labels but changes the simulated operating conditions. This test measures whether the QF-LOG workflow remains stable under shifted generation settings.

Cross-generator performance is reported with the same metrics used for the Generator-A holdout test. The macro F1-score change from Generator-A to Generator-B summarizes performance reduction. Generator-B results are also examined by operating profile and observation window to identify shifted settings with greater classification difficulty.

Forensic interpretation uses class-wise feature behavior, feature contribution results, and confusion-matrix patterns. Increased \texttt{qber} is examined for partial intercept-resend, reduced \texttt{photon\_count} for fiber-tap, and combined count reduction with latency change for detector-blind. The analysis focuses on simulated log overlap and does not treat model output as evidence of attacks on deployed QKD hardware.


\section{Results}
\label{sec:results}

\subsection{Internal Performance and Feature Dependence}
\label{subsec:internal_feature_results}

Table~\ref{tab:generator_a_performance} reports the Generator-A holdout and repeated-validation results. All full-feature learning models outperformed the transparent two-feature comparator. XGBoost achieved the highest holdout macro F1-score of \(0.8243\), followed closely by random forest at \(0.8242\) and logistic regression at \(0.8211\). Repeated validation showed the same pattern, with random forest reaching \(0.8245 \pm 0.0003\) and XGBoost reaching \(0.8243 \pm 0.0003\). McNemar testing showed that each full-feature learning method significantly outperformed the transparent two-feature comparator ($p < 0.001$), while random forest and XGBoost were not significantly different on Generator-A ($p = 0.7457$).

\begin{table}[t]
\centering
\caption{Generator-A internal evaluation results. Panel A reports fixed holdout performance. Panel B reports repeated-validation macro F1-score as mean $\pm$ standard deviation.}
\label{tab:generator_a_performance}
\footnotesize
\renewcommand{\arraystretch}{1.00}
\textbf{Panel A: Fixed holdout evaluation}

\vspace{2pt}
\begin{tabular}{@{}lccccc@{}}
\toprule
\textbf{Method} & \textbf{Accuracy} & \textbf{Macro F1} & \textbf{MCC} & \textbf{ROC-AUC} & \textbf{PR-AUC} \\
\midrule
Baseline             & 0.7461 & 0.7524 & 0.6634 & --     & --     \\
Logistic regression  & 0.8197 & 0.8211 & 0.7600 & 0.9585 & 0.8819 \\
Random forest        & 0.8227 & 0.8242 & 0.7648 & 0.9595 & 0.8838 \\
XGBoost              & 0.8228 & 0.8243 & 0.7648 & 0.9595 & 0.8835 \\
\bottomrule
\end{tabular}

\vspace{4pt}
\textbf{Panel B: Repeated-validation macro F1-score}

\vspace{2pt}
\begin{tabular}{@{}lc@{}}
\toprule
\textbf{Method} & \textbf{Macro F1 Mean $\pm$ SD} \\
\midrule
Baseline             & \(0.7527 \pm 0.0003\) \\
Logistic regression  & \(0.8212 \pm 0.0004\) \\
Random forest        & \(0.8245 \pm 0.0003\) \\
XGBoost              & \(0.8243 \pm 0.0003\) \\
\bottomrule
\end{tabular}
\end{table}

 Table~\ref{tab:feature_diagnostic_summary} summarizes feature dependence
within the configured generator.

\begin{table}[htbp]
\centering
\caption{Summary of feature diagnostic results across logistic regression, random forest, and XGBoost.}
\label{tab:feature_diagnostic_summary}
\footnotesize
\setlength{\tabcolsep}{9pt}
\renewcommand{\arraystretch}{0.90}
\begin{tabularx}{\linewidth}{@{}lccc@{}}
\toprule
\textbf{Feature} &
\makecell{\textbf{Permutation}\\\textbf{Importance Range}} &
\makecell{\textbf{F1 Decrease}\\\textbf{When Removed}} &
\makecell{\textbf{Best Single-}\\\textbf{Feature F1}} \\
\midrule
QBER & 0.2950--0.3047 & 0.0806--0.0810 & 0.5014 \\
Latency (ms) & 0.2422--0.2624 & 0.0547--0.0551 & 0.5835 \\
photon count & 0.0901--0.1479 & 0.0170--0.0173 & 0.5485 \\
abort flag & 0.0024--0.0096 & 0.0006--0.0016 & 0.2943 \\
\bottomrule
\end{tabularx}
\end{table}

\subsection{Class-Level and Cross-Generator Assessment}
\label{subsec:cross_generator_results}

\begin{figure}[t]
\centering
\begin{minipage}[t]{0.49\textwidth}
    \centering
    \includegraphics[width=\linewidth]
    {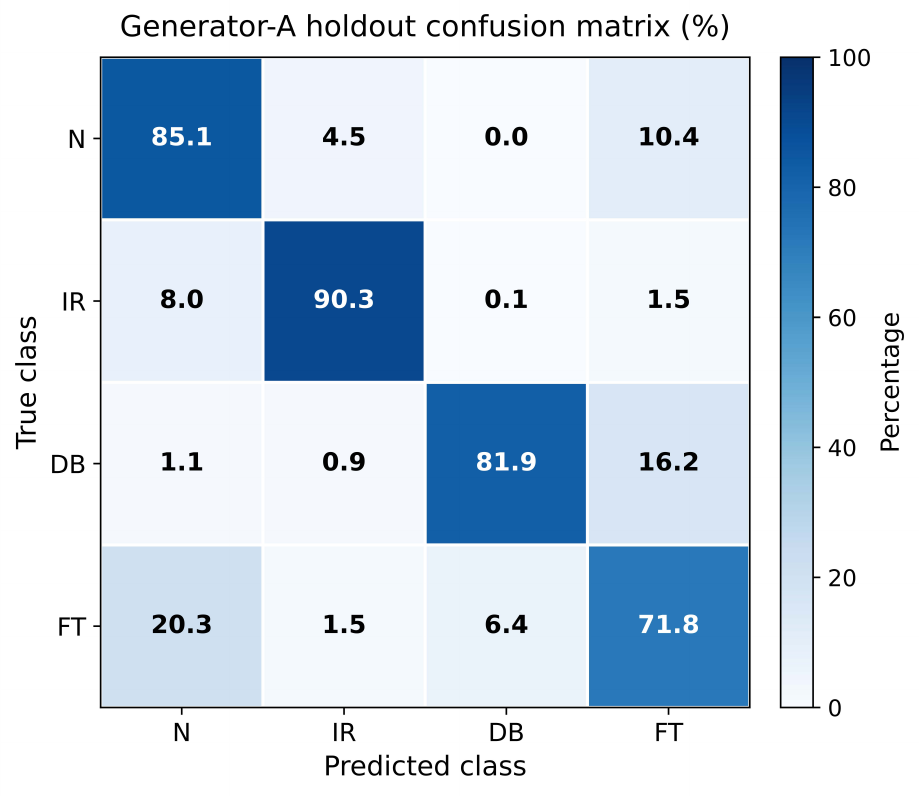}
    \vspace{-2mm}

    {\small (a) Generator-A holdout}
\end{minipage}
\hfill
\begin{minipage}[t]{0.49\textwidth}
    \centering
    \includegraphics[width=\linewidth]
    {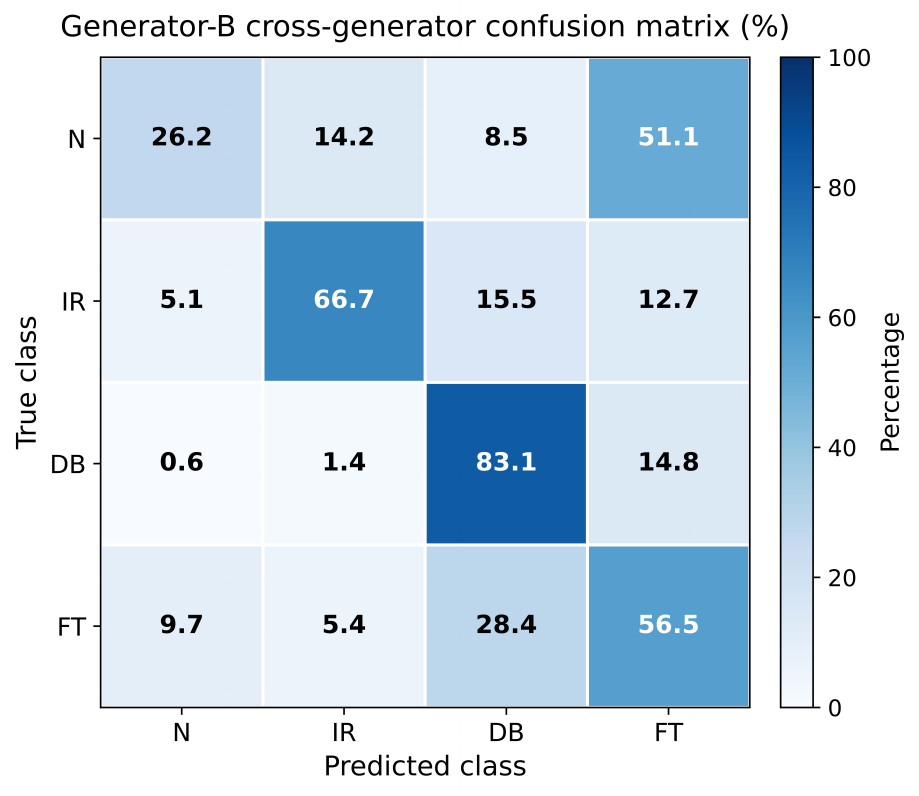}
    \vspace{-2mm}

    {\small (b) Generator-B cross-generator}
\end{minipage}
\vspace{-1mm}
\caption{Row-normalized Random Forest confusion matrices for
(a) the Generator-A fixed holdout evaluation and
(b) the Generator-B cross-generator evaluation.
Rows denote true classes and columns denote predicted classes.
N: normal; IR: partial intercept-resend; DB: detector-blind;
FT: fiber-tap. Values are percentages.}
\label{fig:rf_confusion}
\end{figure}

Random forest is used for class-level interpretation because its
Generator-A performance is effectively equal to that of XGBoost.
Figure~\ref{fig:rf_confusion} shows that Generator-A errors are
concentrated between normal and fiber-tap: 20.3\% of fiber-tap
records are predicted as normal, while 10.4\% of normal records are
predicted as fiber-tap. Under Generator-B, 51.1\% of normal records
are predicted as fiber-tap, and 28.4\% of fiber-tap records are
predicted as detector-blind. These errors show greater class overlap
under the shifted simulated operating conditions.

\begin{table}[t]
\centering
\caption{Random Forest class-level results on Generator-A and
Generator-B. P, R, and F1 denote precision, recall, and F1-score.
FPR denotes the proportion of normal records incorrectly predicted
as the corresponding attack-related class.}
\label{tab:rf_class_metrics}
\small
\setlength{\tabcolsep}{3.2pt}
\begin{tabular}{@{}lcccccccc@{}}
\toprule
& \multicolumn{4}{c}{Generator-A}
& \multicolumn{4}{c}{Generator-B} \\
\cmidrule(lr){2-5}\cmidrule(lr){6-9}
Class
& P & R & F1 & FPR
& P & R & F1 & FPR \\
\midrule
Normal
& 0.7433 & 0.8510 & 0.7935 & -- 
& 0.6305 & 0.2624 & 0.3706 & -- \\

Partial IR
& 0.9295 & 0.9031 & 0.9161 & 0.0446
& 0.7602 & 0.6670 & 0.7105 & 0.1423 \\

Detector-blind
& 0.9260 & 0.8189 & 0.8691 & 0.0001
& 0.6134 & 0.8314 & 0.7059 & 0.0845 \\

Fiber-tap
& 0.7185 & 0.7180 & 0.7182 & 0.1042
& 0.4181 & 0.5648 & 0.4805 & 0.5108 \\
\bottomrule
\end{tabular}
\end{table}

Table~\ref{tab:rf_class_metrics} reports the corresponding class-level
metrics. On Generator-A, partial intercept-resend achieved the highest
class F1-score of 0.9161, while fiber-tap achieved 0.7182. Under
Generator-B, the normal-class F1-score decreased to 0.3706, and the
normal-to-fiber-tap false-positive rate increased from 0.1042 to
0.5108.

Table~\ref{tab:generator_b_performance} reports performance when methods fitted on Generator-A were applied directly to Generator-B without retraining or tuning. All learning-based methods showed substantial macro F1-score reduction. Logistic regression decreased from \(0.8211\) to \(0.5645\), random forest decreased from \(0.8242\) to \(0.5669\), and XGBoost decreased from \(0.8243\) to \(0.5665\). The transparent baseline also decreased from \(0.7524\) to \(0.5083\).

McNemar testing showed that each full-feature learning method remained significantly better than the transparent two-feature comparator on Generator-B (\(p<0.001\)). Random forest and XGBoost produced statistically different predictions, but their macro F1-score difference was below \(0.001\). This difference is not treated as a meaningful practical advantage. Overall, the cross-generator results show that strong Generator-A performance does not establish robustness under shifted simulated operating conditions.

\begin{table}[htbp]
\centering
\caption{Generator-B performance and macro F1-score change relative to the Generator-A holdout evaluation.}
\label{tab:generator_b_performance}
\scriptsize
\setlength{\tabcolsep}{7pt}
\renewcommand{\arraystretch}{1.35}
\begin{tabularx}{\linewidth}{@{}lccccccc@{}}
\toprule
\textbf{Method} &
\makecell{\textbf{A}\\\textbf{F1}} &
\makecell{\textbf{B}\\\textbf{Acc.}} &
\makecell{\textbf{B}\\\textbf{F1}} &
\makecell{\textbf{\(\Delta\)}\\\textbf{F1}} &
\makecell{\textbf{B}\\\textbf{MCC}} &
\makecell{\textbf{B}\\\textbf{AUC}} &
\makecell{\textbf{B}\\\textbf{AP}} \\
\midrule
Baseline & 0.7524 & 0.5297 & 0.5083 & -0.2441 & 0.3858 & -- & -- \\
Logistic regression & 0.8211 & 0.5844 & 0.5645 & -0.2566 & 0.4580 & 0.8547 & 0.6708 \\
Random forest & 0.8242 & 0.5814 & 0.5669 & -0.2574 & 0.4534 & 0.8507 & 0.6661 \\
XGBoost & 0.8243 & 0.5810 & 0.5665 & -0.2578 & 0.4529 & 0.8504 & 0.6646 \\
\bottomrule
\end{tabularx}
\end{table}

\subsection{Observable Shift and Simulated Operating-Profile Analysis}
\label{subsec:shift_analysis}

Generator-B introduced measurable shifts relative to Generator-A. The largest change was in \texttt{photon\_count}, which decreased by \(152.95\) counts on average, with an SMD of \(-0.5879\). Latency increased by \(0.226\) ms, QBER increased by \(0.00329\), and \texttt{abort\_flag} decreased by \(0.05641\). These shifts increased class overlap and contributed to the lower cross-generator performance.

\begin{table}[t]
\centering
\caption{Generator-B macro F1-score by simulated operating profile.}
\label{tab:hardware_profile_results}
\footnotesize
\setlength{\tabcolsep}{5pt}
\renewcommand{\arraystretch}{1.06}
\begin{tabular}{@{}lccc@{}}
\toprule
\textbf{Simulated Operating Profile} & \textbf{LR} & \textbf{RF} & \textbf{XGB} \\
\midrule
Longer fiber link & 0.4360 & 0.4291 & 0.4302 \\
Low-efficiency detector & 0.5267 & 0.5318 & 0.5318 \\
Shifted standard & 0.6471 & 0.6518 & 0.6509 \\
\bottomrule
\end{tabular}
\end{table}

Table~\ref{tab:hardware_profile_results} shows that the reduction was strongest under the longer-fiber-link profile, where all learning-based models stayed below \(0.44\) macro F1-score. The shifted-standard profile was less difficult, with macro F1-scores near \(0.65\). Performance changed little across observation windows, which indicates that the main difficulty came from shifted simulated channel and detector settings.

\subsection{Forensic Interpretation}
\label{subsec:forensic_interpretation}

The observed class-wise patterns support structured forensic interpretation under controlled simulation settings. Partial intercept-resend is mainly associated with increased QBER while its mean photon count remains close to normal operation. Fiber-tap is mainly associated with reduced photon count while its mean QBER remains close to normal. The simulated detector-blind condition produces the strongest photon-count reduction together with higher latency and abort rate.

These findings also define the limit of interpretation. Within Generator-A, the full-feature representation outperforms the
transparent two-feature comparator, while no single stored feature
reproduces full-model performance. Under Generator-B, the same methods show substantial performance reduction as simulated operating settings change. The proposed framework therefore supports forensic analysis of generated QKD operational-log patterns, but it does not establish deterministic attack rules or performance on hardware-generated QKD logs.


\section{Discussion}
\label{sec:discussion}

On Generator-A, the full-feature learning models achieved macro F1-scores from 0.8211 to 0.8243, compared with 0.7524 for the transparent two-feature comparator. Because the comparator uses only QBER and photon count, this difference reflects both model form and access to latency and abort flag. Random forest and XGBoost were not significantly different.

The feature analyses provide insight into the learned decision process. Across ablation and permutation analyses, QBER produced the largest macro-F1 decrease under permutation and ablation within the configured Generator-A dependency structure, followed by latency in milliseconds and photon count. The abort flag contributed comparatively little; no single stored attribute reproduced the complete feature set’s performance, although the stored attributes are mathematically dependent by construction, indicating that classification depends on the combined behavior of multiple observable attributes rather than a single dominant indicator. Class-level results showed strongest recognition for partial intercept-resend and greater ambiguity for fiber-tap, mainly because reduced photon count can overlap with normal channel variation.

The central finding of the study is the substantial performance reduction observed under cross-generator assessment. Macro F1-score decreased from approximately \(0.82\) on Generator-A to approximately \(0.56\) on Generator-B for all learning-based methods. The largest degradation occurred under the longer-fiber operating profile, while observation-window effects were comparatively small. These results show that strong performance on one synthetic generator does not by itself demonstrate robustness under changed simulated operating conditions. Therefore, the framework should be viewed as a simulation-based forensic evaluation approach rather than evidence of hardware-validated or deployment-ready QKD attack detection.


\section{Limitations}
\label{sec:limitations}

QF-LOG has five main limitations. First, it uses only four observable features and does not include detector timing, optical power, calibration, or diagnostic records. Second, both datasets are physics-inspired simulations rather than measurements from deployed QKD hardware, so the results demonstrate simulation-based feasibility only. Third, the study covers three simulated attack conditions and excludes other threats such as time-shift, phase-remapping, photon-number-splitting, and wavelength-dependent attacks \cite{scarani2009security}. Fourth, each record is classified independently, without modelling temporal attack behavior across consecutive frames. Finally, Generator-B represents only one shifted synthetic generator; therefore, further evaluation with additional generators and hardware-generated QKD logs is required.

\section{Future Work}
\label{sec:futurework}

Future work should evaluate additional generators and hardware-generated logs, extend QF-LOG with detector timing, optical power, calibration, and controller diagnostics, and study sequence-level analysis for gradual drift and persistent abnormal behavior.

\section{Conclusion}
\label{sec:conclusion}

This study presented QF-LOG, a simulation-based framework for forensic classification of QKD operational-log behavior using QBER, photon count, latency in milliseconds, and abort flag. The framework evaluated normal, partial intercept-resend, detector-blind, and fiber-tap conditions under controlled physics-inspired simulation settings.

On Generator-A, the full-feature learning models achieved macro F1-scores from 0.8211 to 0.8243, compared with 0.7524 for the transparent two-feature comparator. Feature analysis showed that QBER, latency in milliseconds, and photon count jointly supported classification, while no single feature reproduced the full-model performance. Cross-generator testing produced the central finding. When models trained on Generator-A were tested directly on shifted Generator-B data, macro F1-scores decreased to approximately 0.56--0.57. This result shows that stable internal performance on one synthetic generator does not establish reliability under changed simulated QKD operating conditions.

QF-LOG therefore provides an interpretable foundation for studying forensic classification from QKD operational logs. Further evaluation across broader simulated settings and hardware-generated logs is required before practical forensic deployment can be considered.

\section*{Disclosure of Interests}
The authors have no competing interests to declare that are relevant
to the content of this article.


\bibliographystyle{splncs04}
\bibliography{refs}

@article{gisin2002quantum,
  title        = {Quantum cryptography},
  author       = {Gisin, Nicolas and Ribordy, Gr{\'e}goire and Tittel, Wolfgang and Zbinden, Hugo},
  journal      = {Reviews of Modern Physics},
  volume       = {74},
  number       = {1},
  pages        = {145--195},
  year         = {2002}
}

@article{scarani2009security,
  title        = {The security of practical quantum key distribution},
  author       = {Scarani, Valerio and Bechmann-Pasquinucci, Helle and Cerf, Nicolas J and D{\'u}{\v{s}}ek, Miloslav and L{\"u}tkenhaus, Norbert and Peev, Momtchil},
  journal      = {Reviews of Modern Physics},
  volume       = {81},
  number       = {3},
  pages        = {1301--1350},
  year         = {2009}
}

@article{diamanti2016practical,
  title        = {Practical challenges in quantum key distribution},
  author       = {Diamanti, Eleni and Lo, Hoi-Kwong and Qi, Bing and Yuan, Zhen},
  journal      = {npj Quantum Information},
  volume       = {2},
  pages        = {16025},
  year         = {2016}
}

@article{lo2014secure,
  title        = {Secure quantum key distribution},
  author       = {Lo, Hoi-Kwong and Curty, Marcos and Tamaki, Kiyoshi},
  journal      = {Nature Photonics},
  volume       = {8},
  number       = {8},
  pages        = {595--604},
  year         = {2014}
}

@article{pirandola2020advances,
  title        = {Advances in quantum cryptography},
  author       = {Pirandola, Stefano and Andersen, Ulrik Lund and Banchi, Leonardo and others},
  journal      = {Advances in Optics and Photonics},
  volume       = {12},
  number       = {4},
  pages        = {1012--1236},
  year         = {2020}
}

@article{lydersen2010hacking,
  title        = {Hacking commercial quantum cryptography systems by tailored bright illumination},
  author       = {Lydersen, Lars and Wiechers, Carlos and Wittmann, Christoffer and Elser, Dirk and Skaar, Johannes and Makarov, Vadim},
  journal      = {Nature Photonics},
  volume       = {4},
  number       = {10},
  pages        = {686--689},
  year         = {2010}
}

@article{bennett1984bb84,
  title        = {Quantum cryptography: Public key distribution and coin tossing},
  author       = {Bennett, Charles H and Brassard, Gilles},
  journal      = {Proceedings of IEEE International Conference on Computers, Systems and Signal Processing},
  pages        = {175--179},
  year         = {1984}
}

@article{makarov2006,
  title        = {Effects of detector efficiency mismatch on security of quantum cryptosystems},
  author       = {Makarov, Vadim and Hjelme, Dag R.},
  journal      = {Journal of Modern Optics},
  volume       = {52},
  pages        = {691--705},
  year         = {2006}
}

@inproceedings{qi2007timeshift,
  title        = {Time-shift attack in practical quantum cryptosystems},
  author       = {Qi, Bing and Fung, Chi-Hang Fred and Lo, Hoi-Kwong and Ma, Xiongfeng},
  booktitle    = {Quantum Electronics and Laser Science Conference},
  pages        = {1--2},
  year         = {2007}
}

@article{gerhardt2011blinding,
  title        = {Full-field implementation of a detector-blinding attack},
  author       = {Gerhardt, Ilja and Liu, Qin and others},
  journal      = {Nature Communications},
  volume       = {2},
  pages        = {349},
  year         = {2011}
}

@article{hwang2003decoy,
  title        = {Quantum key distribution with high loss: Decoy states},
  author       = {Hwang, Won-Young},
  journal      = {Physical Review Letters},
  volume       = {91},
  pages        = {057901},
  year         = {2003}
}

@article{lo2012mdi,
  title        = {Measurement-device-independent quantum key distribution},
  author       = {Lo, Hoi-Kwong and Curty, Marcos and Qi, Bing},
  journal      = {Physical Review Letters},
  volume       = {108},
  pages        = {130503},
  year         = {2012}
}

@article{yuan2011monitoring,
  title        = {Robust detection monitoring for practical QKD},
  author       = {Yuan, Zheng and Shields, Andrew},
  journal      = {Applied Physics Letters},
  volume       = {98},
  number       = {231104},
  year         = {2011}
}

@article{breiman2001random,
  title        = {Random forests},
  author       = {Breiman, Leo},
  journal      = {Machine Learning},
  volume       = {45},
  number       = {1},
  pages        = {5--32},
  year         = {2001}
}

@inproceedings{chen2016pxgboost,
  title        = {XGBoost: A scalable tree boosting system},
  author       = {Chen, Tianqi and Guestrin, Carlos},
  booktitle    = {Proc. ACM SIGKDD},
  pages        = {785--794},
  year         = {2016}
}

@article{Ma2005PracticalDecoy,
  author  = {Ma, Xiongfeng and Qi, Bing and Zhao, Yi and Lo, Hoi-Kwong},
  title   = {Practical Decoy State for Quantum Key Distribution},
  journal = {Physical Review A},
  volume  = {72},
  number  = {1},
  pages   = {012326},
  year    = {2005},
  doi     = {10.1103/PhysRevA.72.012326}
}

@article{mcnemar1947note,
  author  = {McNemar, Quinn},
  title   = {Note on the Sampling Error of the Difference Between Correlated Proportions or Percentages},
  journal = {Psychometrika},
  volume  = {12},
  number  = {2},
  pages   = {153--157},
  year    = {1947},
  doi     = {10.1007/BF02295996}
}

@article{gobby2004qkd,
  author  = {Gobby, C. and Yuan, Z. L. and Shields, A. J.},
  title   = {Quantum Key Distribution over 122 km of Standard Telecom Fiber},
  journal = {Applied Physics Letters},
  volume  = {84},
  number  = {19},
  pages   = {3762--3764},
  year    = {2004},
  doi     = {10.1063/1.1738173}
}

@misc{das2025qkdforensicdataset,
  author    = {Das, Madhab Chandra},
  title     = {{QKD-Forensic-Dataset}: Initial Dataset Release},
  year      = {2025},
  publisher = {Zenodo},
  version   = {v1.0.0},
  doi       = {10.5281/zenodo.17773084},
  note      = {Dataset}
}

\end{document}